\documentclass[letterpaper, 10 pt, conference]{ieeeconf}  

\IEEEoverridecommandlockouts                              
\usepackage{graphics} 
\usepackage{epsfig} 
\usepackage{mathptmx} 
\usepackage{times} 
\usepackage{amsmath} 
\usepackage{amssymb}  
\usepackage{amsthm}
\usepackage{algorithm}
\usepackage{enumitem}
\usepackage[noend]{algpseudocode}

\theoremstyle{plain}
\newtheorem{theorem}{Theorem}
\newtheorem{lemma}{Lemma}
\newtheorem{corollary}{Corollary}
\newcommand{\Mod}[1]{\ (\mathrm{mod}\ #1)}

\title{\LARGE \bf
Learning Optimal Stable Matches in Decentralized Markets with Unknown Preferences
}

\author{Vade Shah, Bryce L. Ferguson, and Jason R. Marden 
\thanks{This work is supported by ONR grant \#N00014-20-1-2359 and AFOSR grants \#FA9550-20-1-0054 and \#FA9550-21-1-0203.}
\thanks{V. Shah ({\tt\small vade@ucsb.edu}), B. L. Ferguson, and J. R. Marden are with the Department of Electrical and Computer Engineering at the University of California, Santa Barbara, CA.}%
}

\begin{document}

\maketitle
\thispagestyle{empty}
\pagestyle{empty}

\begin{abstract} Matching algorithms have demonstrated great success in several practical applications, but they often require centralized coordination and plentiful information. In many modern online marketplaces, agents must independently seek out and match with another using little to no information. For these kinds of settings, can we design decentralized, limited-information matching algorithms that preserve the desirable properties of standard centralized techniques? In this work, we constructively answer this question in the affirmative. We model a two-sided matching market as a game consisting of two disjoint sets of agents, referred to as proposers and acceptors, each of whom seeks to match with their most preferable partner on the opposite side of the market. However, each proposer has no knowledge of their own preferences, so they must learn their preferences while forming matches in the market. We present a simple online learning rule that guarantees a strong notion of probabilistic convergence to the welfare-maximizing equilibrium of the game, referred to as the proposer-optimal stable match. To the best of our knowledge, this represents the first completely decoupled, communication-free algorithm that guarantees probabilistic convergence to an optimal stable match, irrespective of the structure of the matching market.
\end{abstract}


\section{Introduction}

How should partnerships be formed?  Although ostensibly simple, the task of \textit{matching} members of two distinct groups is inherently combinatorially challenging. The \textit{two-sided matching} problem has thus motivated decades of research by computer scientists, economists, and engineers alike that has fundamentally transformed the landscape of our social and engineered systems \cite{abdulkadiroglu2013matching, azevedo2016matching, knuth1997stable, irving1985efficient}.

Consider the residency matching process, a prototypical example of a two-sided matching problem in which residents and hospitals must be partnered with one another. In practice, residents typically rank order the hospitals, as do hospitals rank order the residents. A central matchmaker would ideally use these orderings to identify a desirable \textit{match} between residents and hospitals; in particular, one would hope to find a \textit{stable match}, characterized by the property that no resident and hospital would mutually prefer one another to their partners assigned by the matchmaker.

\begin{figure}
\centering
\includegraphics[width=\linewidth]{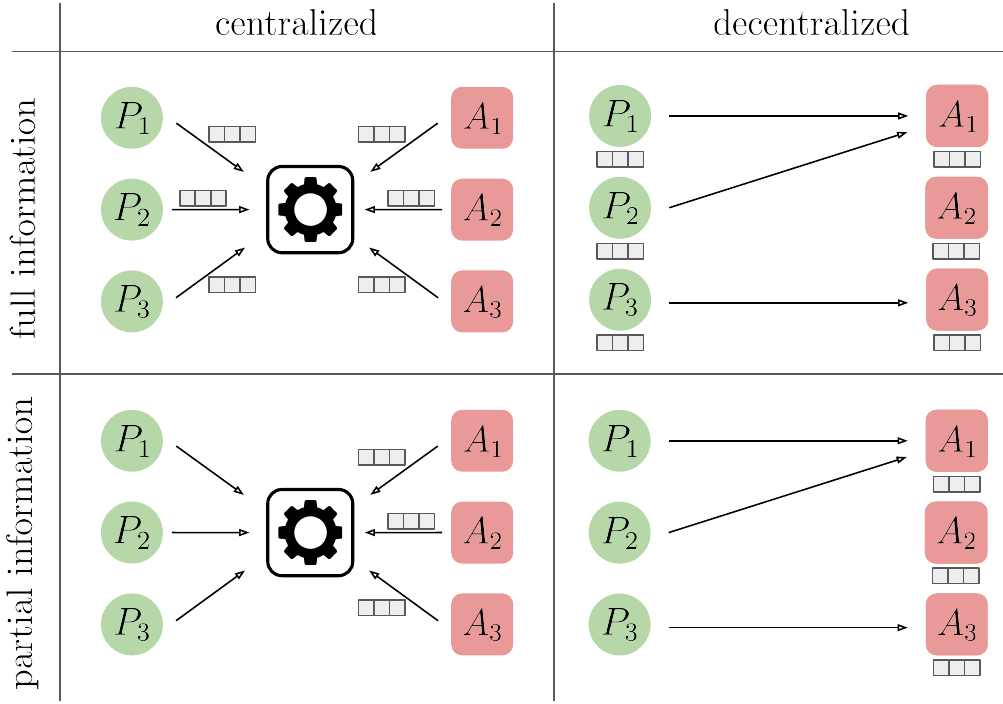}
\caption{Two-sided matching market models with full (top) and partial information (bottom) in centralized (left) and decentralized environments (right). In a full information model, proposers know their own preferences completely, but in a partial information model, they do not. In a centralized environment, agents provide their preferences (or a subset thereof) to a central algorithm that assigns a match, but in a decentralized environment, proposers directly propose to acceptors to form a match.}
\label{fig:market_comparison}
\end{figure}

Celebrated work by Gale and Shapley \cite{gale1962college} demonstrates the existence of such a stable match in every two-sided market. Given a list of the preference orderings of every agent, the Gale-Shapley algorithm (GS) identifies the best possible stable match for all agents on one side of the market (Figure \ref{fig:market_comparison}, top-left). GS and its variants have been implemented in residency admissions and organ donation programs, guaranteeing strong notions of stability and optimality to the arrangements made by these institutions \cite{chen2006school, abdulkadirouglu2005new}.

However, GS is not applicable in all two-sided matching settings. In many scenarios, there is no central clearinghouse that facilitates the matching process, so an algorithm like GS cannot be used. In these kinds of markets (e.g., most labor markets), there is typically `active' set of agents, the \textit{proposers} (firms), who repeatedly make proposals to a `passive' set of agents, the \textit{acceptors} (workers), who accept or reject these proposals (Figure \ref{fig:market_comparison}, top-right). Motivated by these kinds of decentralized environments, scholars have extensively studied the stability and quality of the matches that may result when agents match independently \cite{cheng2016stable, diamantoudi2015decentralized, niederle2009decentralized}.

GS and similar algorithms are well-suited for situations where users report comprehensive preference orderings, but there are many applications where such requirements are prohibitive (Figure \ref{fig:market_comparison}, bottom-left). Generating rankings may become difficult because of issues related to size, a lack of information, or both. For large practical markets like the National Resident Matching Program \cite{chen2006school, abdulkadirouglu2005new, peranson1995nrmp}, several algorithms have been developed to handle the challenges associated with shortened or nonexistent preference orderings \cite{fershtman2017pandora, su2022optimizing}.

This work focuses on markets that face a lack of both centrality and information. A particular motivation is the modern online marketplace, where massive numbers of consumers and producers seek to exchange resources without information on one another or guidance from a central entity (Figure \ref{fig:market_comparison}, bottom-right). Agents repeatedly interact with one another and receive only limited information on the quality of their matched partners through the form of some representative reward. In these kinds of settings, there is a growing need for decentralized \textit{learning rules}, policies that allow agents to learn from their past rewards so that their future behavior leads towards collectively desirable stable matches.

Several recent papers have made significant progress towards this goal for the setting where agents on the proposing side of the market do not know their own preferences \cite{liu2014stable, echenique2024experimental}. Many draw from the multi-armed bandits literature to construct algorithms that provide bounds on the regret experienced by each proposer \cite{maheshwari2022decentralized, cen2022regret, jagadeesan2021learning, liu2020competing, pokharel2023converging}, while others use more general probabilistic techniques that yield similar results \cite{min2022learn, ashlagi2020clearing}. However, these learning rules typically require some form of online communication or coordination between proposers or a central entity, and they also impose restrictive assumptions on the structure of the agents' preferences. Furthermore, the guarantees on the stability or quality of the resulting matches when using these rules are limited.

In this paper, we present a novel algorithm that guarantees stability in decentralized partial-information settings. In particular, we design a \textbf{learning rule} for the proposers to independently follow that provably guarantees a strong notion of probabilistic convergence to their \textbf{optimal stable match} even when they do not know their own preferences. The learning rule thus recovers the classic result of Gale and Shapley for the case of one-sided unknown preferences.

Our learning rule draws from the literature on \textit{learning in games}, a well-established subfield of game theory concerned with the design of online learning algorithms \cite{young1993evolution, young2009learning}. For several types of games, scholars have designed decentralized learning rules that guarantee convergence to desirable configurations such as correlated, Nash, and even welfare-maximizing equilibria \cite{marden2014achieving, hart2000simple, marden2009payoff, pradelski2012learning, leslie2011equilibrium}. Inspired by these results, we first formulate the decentralized two-sided matching market as a game in Section \ref{section:model}. Then, in Section \ref{section:results}, we assert the existence of a novel learning rule that guarantees convergence to the proposer-optimal stable match. We describe and simulate this rule in Sections \ref{section:rule} and \ref{section:simulation}, respectively, and prove its convergence properties in Section \ref{section:proofs} by considerably extending upon existing results from the literature on matching theory.

\section{Model}\label{section:model}

\subsection{Market Model}

\begin{figure}
    \centering
    \includegraphics[width=\linewidth]{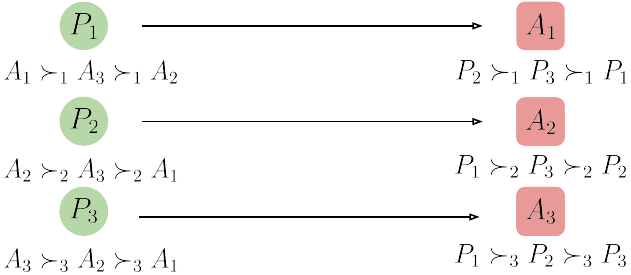}
    \caption{A market with three proposers (left) and three acceptors (right). Each agents' preference ordering is shown as a list beneath their icon. The arrows indicate to whom each proposer proposes; this configuration corresponds to the proposer-optimal stable match in which every proposer is matched with their favorite acceptor.}
    \label{fig:posm}
\end{figure}

The two-sided market consists of two groups: a set of \textit{proposers}, denoted by $\mathbf{P} \triangleq \{ P_1, P_2, \dots, P_n \}$, and a set of \textit{acceptors}, denoted by $\mathbf{A} \triangleq \{ A_1, A_2, \dots, A_m \}$. Each agent has a complete, transitive, and strict \textit{preference ordering}, also referred to as \textit{preferences}, over the agents in the other group. We represent the preference ordering of proposer $P_i$ as a function\footnote{We choose this interval for simplicity, but any interval (open or closed) is acceptable} $O_{P_i} : \mathbf{A} \cup \{\emptyset\} \to [0, 1]$, where we say that $P_i$ \textit{prefers} $A_{j_1}$ to $A_{j_2}$ if $O_{P_i}(A_{j_1}) > O_{P_i}(A_{j_2})$, which we shorthand as $A_{j_1} \succ_i A_{j_2}$; for completeness, we assume $O_{P_i}(A_j) > O_{P_i}(\emptyset) = 0$ for all $j \in \{1, \dots m\}$. We define similarly the preference ordering $O_{A_j}$ of acceptor $A_j$, and as a slight abuse of notation, we use the same shorthand $\succ_j$ to describe the preferences of acceptor $A_j$. The collection of all preference orderings is given by $\mathbf{O} \triangleq \{O_{P_1}, \dots, O_{P_n}, O_{A_1}, \dots, O_{A_m} \}$, so that a particular instance of the matching market game is fully parametrized by the tuple $\mathbf{M} = (\mathbf{P}, \mathbf{A}, \mathbf{O})$. A simple market is depicted in Figure \ref{fig:posm} with $n = m = 3$ proposers and acceptors.

\subsection{Game Model}

We study interactions between agents in the two-sided matching market using a game theoretic framework. A market $\mathbf{M}$ defines a game played by the set of proposers, where each proposer $P_i \in \mathbf{P}$ selects an action $a_i \in \mathbf{A} \cup \{ \emptyset \}$. If $a_i = A_j$, we say that $P_i$ \textit{proposes} to $A_j$, and if $a_i = \emptyset$, we say that $P_i$ chooses to \textit{remain single}. The full action profile is denoted by $a \triangleq \{ a_1, \dots, a_n \}$. After the proposers propose, each acceptor $A_j$ \textit{accepts} their most preferred proposer in $\{ P_i \in \mathbf{P} : a_i = A_j \}$  (where `most preferred' is specified by $O_{A_j}$), and rejects all other proposers that have proposed to them. Since we assume that acceptors behave determinstically by accepting their most preferred proposal from those they receive, we do not regard them as players in the game.

We denote the outcome of these acceptances and rejections as a \textit{match} $\mu(a) \triangleq \{ \mu_{P_1}(a), \dots, \mu_{P_n}(a), \mu_{A_1}(a), \dots, \mu_{A_m}(a) \}$, where $\mu_{P_i}(a) \in \mathbf{A} \cup \{ \emptyset \}$ is the match (or lack thereof) of proposer $P_i$ given the action profile $a$, and $\mu_{A_j}(a) \in \mathbf{P} \cup \{ \emptyset \}$ denotes the outcome for $A_j$. If $\mu_{P_i}(a) = A_j$ then $\mu_{A_j}(a) = P_i$, and we say that the pair $(P_i, A_j)$ are \textit{matched}. Otherwise, if $\mu_{P_i}(a) = \emptyset$ or $\mu_{A_j}(a) = \emptyset$, then we say that proposer $P_i$ or acceptor $A_j$ is \textit{unmatched}. For ease of notation, we simply write $\mu$, $\mu_{P_i}$ or $\mu_{A_j}$ when the dependence on $a$ is clear.

Using this notation, we can define the \textit{utility} for each proposer $P_i$ as  
\begin{equation*}
    u_i(a) = O_{P_i}(\mu_{P_i}(a)),
\end{equation*}
which is the payoff they receive for matching with their partner. With this definition, the assumption $O_{P_i}(\emptyset) = 0$ ensures that every proposer strictly prefers being matched to being unmatched. 

Let us now discuss the equilibria of the game. Any match $\mu$ in which there exists a proposer-acceptor pair $(P_i, A_j)$ who prefer one another to their partners in $\mu$ (i.e., $O_{P_i}(A_j) > O_{P_i}(\mu_{P_i})$ and $O_{A_j}(P_i) > O_{A_j}(\mu_{A_j})$) is said to be \textit{unstable}, and any match that is not unstable is said to be \textit{stable}. A stable match exists in every two-sided market \cite{gale1962college}. Moreover, there exists a unique \textit{proposer-optimal} stable match (POSM) $\mu^*$, which is characterized by the property that, for any alternate stable match $\mu'$ in the market,
\begin{equation*}
    \mu^*_{P_i} \succeq_i \mu'_{P_i} \qquad \forall i \in \{1, \dots, n\},
\end{equation*}
meaning that every proposer weakly prefers their partner in the POSM to their partner in any other stable match. For example, the proposer optimal stable match in the market shown in Figure \ref{fig:posm} can be described by the list of proposer-acceptor pairs $(P_1, A_1), (P_2, A_2), (P_3, A_3)$.

From this, one can conclude that every pure strategy Nash equilibrium of the game corresponds to a stable matching. More specifically, every Nash equilibrium action profile produces exactly one stable match; however, a stable match may be represented by multiple Nash equilibrium action profiles if $n > m$, since the proposers who are unmatched in a stable match are indifferent between their actions. Furthermore, if one defines the \textit{welfare} of an action profile as the sum of the players' utilities corresponding to that action profile, then it is straightforward to show that any action profile corresponding to the proposer-optimal stable match is a \textit{welfare-maximizing} Nash equilibrium of the game.

\section{Learning Model and Main Result}\label{section:results}

We model the repeated interactions between proposers and acceptors as a series of repeated one-shot matching games that are played in timesteps $t = \{1, 2, \dots\}$. At timestep $t$, every proposer $P_i$ selects their action $a_i^t$, which generates a match $\mu^t$ and a utility $u_i^t(a^t)$ as described in the previous section. We assume that proposer $P_i$ selects $a_i^t$ according to a probability distribution $p_i^t$, which we refer to as their \textit{strategy}. By assumption, their strategy at timestep $t$ can only incorporate their actions and utilities in previous timesteps $1, 2, \dots, \; t-1$. We thus define a \textit{learning rule} as a mechanism that agents use to generate their strategy at each timestep, i.e.,
\begin{equation}\label{eq:learning_rule}
    p_i^t = \Gamma(a_i^1, u_i^1, \dots, a_i^{t-1}, u_i^{t-1}; \mathbf{A} \cup \{\emptyset\}).
\end{equation}
Notice that when using $\Gamma$, a proposer must know only their own action and utility history, as well as their own action space, but nothing else. In particular, they need not have any knowledge of their own true preference ordering, nor do they require information regarding the existence of other proposers or their actions. Such learning rules are described as \textit{completely uncoupled}.

Having defined the relevant concepts, let us now formally restate the central question studied in this paper:

\smallskip

\emph{
    Is there a learning rule that provably guarantees convergence to the proposer-optimal stable match in decentralized markets where proposers do not know their own preferences?
}

\smallskip

We assert the existence of a novel learning rule that answers this question in the affirmative in Theorem 1.

\begin{theorem}
Consider the matching game defined by the market $\mathbf{M}$. For every probability $0 \leq p < 1$, there exists a learning rule $\Gamma$ of the form defined in \eqref{eq:learning_rule} such that, if every proposer selects their action according to $\Gamma$, then for all sufficiently large timesteps $t$, the action profile $a^t$ corresponds to the proposer-optimal stable match $\mu^*$ with probability at least $p$.
\end{theorem}

We take this opportunity to highlight the generality of this result. That is, Theorem 1 holds true without any assumptions on the number of proposers and acceptors or the structure of the preference orderings (i.e., the sequential preference condition, $\alpha$-reducibility), which are necessary for many previous results \cite{maheshwari2022decentralized, cen2022regret, jagadeesan2021learning}. Furthermore, applying the rule is infrastructurally unrestrictive, in the sense that agents must only be informed of $\Gamma$ prior to the game being played; otherwise, there is no need for any platform that facilitates the exchange of information between agents or a central entity for the remainder of the process. This is a necessity in multiple existing methods \cite{liu2020competing, pokharel2023converging} that is not always available in practice.

The specific structure of the learning rule that achieves the desired result in Theorem 1 is discussed is detail in Section \ref{section:rule}. However, we want to emphasize that we do not view the specifics of this learning rule as a key contribution of this paper. Rather, the main contribution resides in the fact that one can achieve the POSM in fully distributed settings with almost no information.

Lastly, we note that the proof of our result requires building upon a number of results on better- and best-reply dynamics in matching games \cite{roth1990random, ackermann2008uncoordinated}, which is quite involved. We argue that both the result and its proof present interesting contributions to matching theory.

\section{Description of Learning Rule}\label{section:rule}

In this section, we define the learning rule that each agent uses to select their action, which follows a similar structure\footnote{In \cite{pradelski2012learning}, the authors present a learning rule that guarantees convergence to welfare-maximizing equilibria. However, these dynamics only apply to games that are interdependent (see  \cite{pradelski2012learning}), which the matching game is not.} to the trial and error learning dynamics considered in \cite{young2009learning, marden2009payoff, pradelski2012learning}. In the learning rule, each proposer maintains a local state variable that impacts their chosen strategy at each timestep. After implementing the given strategy, they receive a utility that in turn influences their state, and the feedback loop between these two processes continues. The specific rules that govern how proposers select their actions and update their states are spelled out in detail below.

In each timestep, proposer $P_i$ generates their strategy $p_i^t$ according to their \textit{state} $x_i \triangleq (m_i, \underline{a}_i, \underline{u}_i)$, where
\begin{itemize}
    \item $m_i \in \{ C, D, W \}$ is the \textit{mood} of proposer $P_i$, which can take one of three values: $C$ (content), $W$ (watchful), or $D$ (discontent);
    \item $\underline{a}_i \in \mathbf{A} \cup \{P_i\}$ is the \textit{baseline action} of proposer $P_i$;
    \item $\underline{u}_i \in \mathbb{R}_{\geq 0}$ is the \textit{baseline utility} of proposer $P_i$.
\end{itemize}

We write the collection of proposer states as $x = (m, \underline{a}, \underline{u})$.  The time-homogeneous processes by which proposers select their actions and update their states are referred to as the \textit{action selection rule} and the \textit{state update rule}, both of which are governed by a common experimentation rate $\epsilon > 0$; we refer to the pair as the \textit{learning rule}. At each timestep, every proposer independently selects and plays their action according to the action selection rule described in Algorithm \ref{alg:action}.

\begin{algorithm}
\caption{Action selection rule}\label{alg:action}
\begin{algorithmic}
\Require $P_i$, $x_i$, $\epsilon$
\If{$m_i = C$}
    \State play baseline action $a_i = \underline{a}_i$ with probability (w.p.) $1 - \epsilon - \epsilon^2$; or
    \State play new action $a_i \in \mathbf{A}$ uniformly at random w.p. $\epsilon$; or
 	\State play new action $a_i = P_i$ w.p. $\epsilon^2$
\ElsIf{$m_i = D$}
	\State play baseline action $a_i = \underline{a}_i = P_i$ w.p. $1 - \epsilon^{1.5}$; or
	\State play new action $a_i \in \mathbf{A}$ uniformly at random w.p. $\epsilon^{1.5}$
\ElsIf{$m_i = W$}
	\State play baseline action $a_i = \underline{a}_i$ w.p. $1$
\EndIf
\end{algorithmic}
\end{algorithm}

After selecting and playing their action $a_i$ and receiving their utility $u_i$, each proposer $P_i$ independently updates their state $x_i$. This update behavior is governed by two functions $F : [0, 1] \to [0, 0.5)$ and $G : [0, 1] \to [0, 0.5)$, where we assume that both functions are strictly monotone decreasing\footnote{The domains and codomains are chosen for simplicity of analysis, but other intervals can be used with appropriate modifications to the model.}. The state update rule is described in Algorithm \ref{alg:state}, where $\Delta u_i \triangleq u_i - \underline{u}_i$.

\begin{algorithm}
\caption{State update rule}\label{alg:state}
\begin{algorithmic}
\Require $P_i$, $x_i$, $u_i$, $\epsilon$, $F$, $G$
\If{$m_i = C$ and $P_i$ did not experiment}
    \State $x_i = \begin{cases}
            (C, \underline{a}_i, u_i) & u_i \geq \underline{u}_i \\
            (W, \underline{a}_i, \underline{u}_i) & u_i < \underline{u}_i
        \end{cases}$
\ElsIf{$m_i = C$ and $P_i$ did experiment}
    \If{$u_i \leq \underline{u}_i$}
        \State $x_i = (C, \underline{a}_i, u_i)$
    \Else{}
        \State $x_i = \begin{cases}
                (C, \underline{a}_i, \underline{u}_i) & \text{w.p. } 1 - \epsilon^{G(\Delta u_i)} \\
                (C, a_i, u_i) & \text{w.p. } \epsilon^{G(\Delta u_i)}
            \end{cases}$
    \EndIf
\ElsIf{$m_i = D$ and $P_i$ did experiment}
	\State $x_i = \begin{cases}
            (D, \underline{a}_i, \underline{u}_i) & \text{w.p. } 1 - \epsilon^{F(u_i)} \\
            (C, a_i, u_i) & \text{w.p. } \epsilon^{F(u_i)}
        \end{cases}$
\ElsIf{$m_i = W$}
	\State $x_i = \begin{cases}
        (C, \underline{a}_i, \underline{u}_i) & u_i \geq \underline{u}_i \\
        (D, P_i, 0) & u_i < \underline{u}_i
    \end{cases}$
\EndIf
\end{algorithmic}
\end{algorithm}

Although lengthy, the rule prescribes qualitatively straightforward behavior. Proposers asynchronously alternate between a content mood, where they successfully match in the market, and a discontent mood, where they are inactive and remain single. Occasionally, the learning rule spurs them to experiment with new actions, which probabilistically induces changes in their own state as well as others' depending on changes in utility. 

\section{Simulation}\label{section:simulation}

\begin{figure}
\centering
\includegraphics[width=\linewidth]{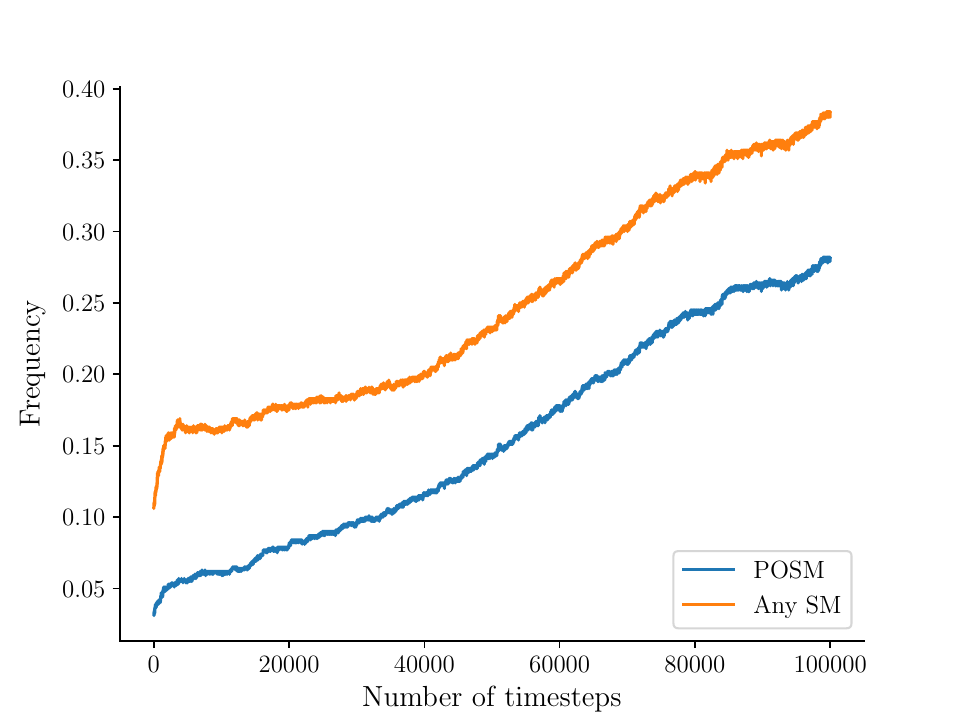}
\caption{When using the learning rule, the empirical frequency of reaching any stable match (SM) and the POSM increases with the number of timesteps when every agent follows the learning rule with parameters $\epsilon = 0.001$, $F(u) = -0.49 \exp (-4u)$, and $G(\Delta u) =  -0.49 \exp (-4 \Delta u)$.}
\label{fig:sim}
\end{figure}

Consider the market with three proposers and three acceptors shown in Figure \ref{fig:posm}. We simulate the performance of our learning rule on this market over 1,000 different initial matches for 100,000 timesteps. The empirical frequency of playing an action corresponding to a stable match or the POSM is shown for each timestep in the right portion of Figure \ref{fig:sim}. Observe that the empirical frequency with which the agents' baseline action corresponds to the POSM increases with the number of periods, as does the frequency with which their action corresponds to any stable match.

We call attention to the fact that the empirical frequency of being at any stable match plateaus within the first 10,000 timesteps, while the frequency of being at the POSM increases. This is an important observation, as it shows that the dynamics successfully incentivize proposers to deviate from suboptimal stable matches. After escaping these configurations, they can then progress towards the POSM.

\section{Conclusion}

In this work, we study a two-sided matching market in which the agents on the proposing side of the market do not know their preferences over the agents on the accepting side. We provide a fully decentralized learning rule that, if every agent follows, guarantees a strong notion of probabilistic convergence to the proposer-optimal stable match. In doing so, we recover the result of Gale and Shapley for decentralized markets with one-sided unknown preferences. 

This work closes a gap in the literature on matching markets with unknown preferences. In particular, to the best of the authors' knowledge, the learning rule presented in this paper is the first to provably guarantee any notion of convergence to a stable match, much less the optimal stable match, and does so without any restrictions on the agents' preference orderings. We view our learning rule as a theoretically grounded starting point for developing more advanced algorithms that increase the rate of convergence while maintaining provable guarantees.

\section{Proof of Theorem}\label{section:proofs}

The proof of Theorem 1 relies on several existing results from the literature on learning in games and two-sided matching markets. First, in Sections \ref{subsection:stochastic} and \ref{subsection:matching}, we present and extend existing results from Markov process theory and matching theory, respectively. Then, we combine these results to complete our proof in Section \ref{subsection:proof}.

\subsection{Stochastic stability}\label{subsection:stochastic}

Let $\mathbf{T}^0$ denote the probability transition matrix of a finite state Markov chain on a state space $\mathbf{X}$. Consider a 'perturbed' version of the process, $\mathbf{T}^\epsilon$, where the 'size' of the perturbation is described by a scalar $\epsilon > 0$. We say that $\mathbf{T}^\epsilon$ is a \textit{regular perturbed Markov process} if the following conditions hold:
\begin{enumerate}
    \item $\mathbf{T}^\epsilon$ is ergodic
    \item $\mathbf{T}^\epsilon$ approaches $\mathbf{T}^0$ at an exponentially smooth rate, meaning that if $\mathbf{T}^\epsilon_{xy} > 0$ for some $x, y \in \mathbf{X}$ and for some $\epsilon > 0$, then $0 < \lim_{\epsilon \to 0^+} \frac{\mathbf{T}_{xy}^\epsilon}{\epsilon^{r(x, y)}} < \infty$.
\end{enumerate}
We refer to the unique positive number $r(x, y)$ as the \textit{resistance} of the transition from state $x$ to state $y$. For completeness, if $\mathbf{T}^\epsilon_{xy} = 0$ for any $\epsilon > 0$, then $r(x, y) = \infty$; similarly, if $\mathbf{T}^0_{xy} > 0$, then $r(x, y) = 0$.

Let $\mathbf{Z} \triangleq \{z_0, z_1, \dots, z_v\}$ denote the set of all recurrence classes of $\mathbf{T}^0$. For each pair of recurrence classes $z_i, z_j$, an $ij$\textit{-path} is a sequence of states $[x] = \{x_1, x_2, \dots, x_k\}$ such that $x_1 \in z_i$ and $x_k \in z_j$. The resistance of the path $[x]$ is the sum of the resistances along each of its transitions, i.e., $r([x]) = r(x_1, x_2) + \dots + r(x_{k-1}, x_k)$. We define the resistance from $z_i$ to $z_j$ as the minimum resistance over all $ij$-paths, which we denote as $\rho(z_i, z_j)$.

Now, consider a graph $G$ with $v$ vertices, each of which is a recurrence class of the unperturbed process $\mathbf{T}^0$, and let the weight of the edge from $z_i$ to $z_j$ be $\rho(z_i, z_j)$. We define a $z_j$\textit{-tree} $T(z_j)$ as a subgraph of $G$ such that $T(z_j)$ has no cycles and for all $z_i \neq z_j$, there exists a unique path from $z_i$ to $z_j$ in $T(z_j)$. The resistance of a $z_j$-tree is the sum of the resistances of the edges that compose it. The \textit{stochastic potential} $\rho(z_j)$ of a node $z_j$ is the minimum resistance over all $z_j$-trees. Lemma \ref{lemma:stochastic_stability} provides a useful criterion for finding the \textit{stochastically stable} states of a regular perturbed Markov process, which are those that are observed with positive probability in the limiting stationary distribution as $\epsilon \to 0$.

\begin{lemma}[Young, 1993 \cite{young1993evolution}]
\label{lemma:stochastic_stability}
	The stochastically stable states of a regular perturbed Markov process $\mathbf{T}^\epsilon$ are precisely the states contained in the recurrence classes of minimum stochastic potential.
\end{lemma}

\subsection{Matching theory}\label{subsection:matching}

Henceforth, we use the notation $\mu(x)$ to denote the partner of agent $x$ in the match $\mu$ for readability. In every two-sided market, the structure of the set of all stable matches can be described by a lattice \cite{knuth1997stable}. Recall that a lattice $\mathbf{L}$ is a partially ordered set with the property that every pair of elements $x, y \in \mathbf{L}$ has a unique supremum and a unique infimum with respect to the set order $\leqslant$, both of which are contained in $\mathbf{L}$. For the lattice of stable matches, we define the order operation $\leqslant$ such that $\mu' \leqslant \mu$ implies that every proposer weakly prefers their partner in $\mu$ to their partner in $\mu'$.

Given a stable match $\mu \neq \mu^*$, we say that $\mu^+$ is a \textit{next-best} stable match with respect to $\mu$ if $\mu^+$ lies directly above $\mu$ on the lattice of stable matches, i.e., there is no intermediate stable match $\mu'$ such that $\mu \leqslant \mu' \leqslant \mu^+$. Let $\{P_{i_1}, \dots, P_{i_r}\} \subset \mathbf{P}$ denote the set of proposers who strictly prefer $\mu^+$ to $\mu$. Assuming that the indices $i_1, \dots, i_r$ are ordered appropriately, we say that $\{P_{i_1}, \dots, P_{i_r}\}$ form a \textit{rotation}, meaning that
\begin{equation*}
\mu(P_{i_k}) = \mu' \left(P_{i_{k \Mod r + 1}} \right), \quad \forall k \in \{1, \dots, r\}.
\end{equation*}
Hence, going between two adjacent stable matches simply requires cyclically shifting a subset of proposers. The concept of a rotation extends to arbitrary pairs of stable matches, but in this text, we exclusively use the term to describe a set of proposers who strictly improve from one stable match to another that lies directly above it on $\mathbf{L}$.

With these ideas in mind, we must design dynamics that can not only reach the lattice of stable matches, but also 'traverse' it in an upwards manner. A useful tool for this task, termed \textit{best response dynamics}, is introduced in \cite{ackermann2008uncoordinated}. Given an initial match $\mu^0$, the best response dynamics produce a sequence of matches in two phases:
\begin{enumerate}
	\item During Phase 1, at each timestep, exactly one matched proposer updates their action to their \textit{best response}, meaning that they propose to their most preferred acceptor who will accept them. This produces a finite sequence of matches $\mu^0, \mu^1, \dots, \mu^k$ with the property that in $\mu^k$, every matched proposer is playing a best response.
	\item During Phase 2, at each timestep, exactly one unmatched proposer updates their action to their best response. This produces a finite sequence of matches $\mu^k, \mu^{k+1}, \dots, \mu^l$.
\end{enumerate}
The best response dynamics satisfy the following property:
\begin{lemma}[Ackermann et al., 2008, \cite{ackermann2008uncoordinated}]
\label{lemma:best_response}
	For every matching market $\mathbf{M}$ with $n$ proposers and $m$ acceptors and every match $\mu^0$, there exists a sequence of at most $2nm$ best responses starting in $\mu^0$ that lead to a match $\mu^l$ that is stable.
\end{lemma}

Consider a stable match $\mu$, and define the match $\mu^{-i}$ as the \textit{near-stable} match resulting from $\mu$ if proposer $P_i$ is made single, i.e.,
\begin{equation*}
\mu^{-i}(P_j) = \begin{cases}
\mu(P_j) & j \neq i \\
P_j & j = i.
\end{cases}
\end{equation*}
The following result establishes a useful property of the best response dynamics initiated at near-stable matches.
\begin{lemma}\label{lemma:lattice_traversal}
	Consider an arbitrary matching market $\mathbf{M}$ and an associated stable match $\mu \neq \mu^*$, and let $\mathbf{P}' = \{P_{i_1}, \dots, P_{i_r} \} \subseteq \mathbf{P}$ be a rotation for some next-best stable match with respect to $\mu$. For every $i \in \{i_1, \dots i_r\}$, every best response sequence starting from the near-stable match $\mu^{-i}$ leads to a stable match $\mu' > \mu$.
\end{lemma}

Before proving this result, we state a useful corollary:

\begin{corollary}\label{cor:lattice_stays}
	Consider an arbitrary matching market $\mathbf{M}$ and an associated stable match $\mu$. For every $i \in \{1, \dots, n\}$, every best response sequence starting from the near-stable match $\mu^{-i}$ leads to a stable match $\mu' \geqslant \mu$.
\end{corollary}

We first prove Lemma \ref{lemma:lattice_traversal}, from which the proof of Corollary \ref{cor:lattice_stays} follows.

\begin{proof}
Let $\mu^+$ denote the next-best stable match from $\mu$ with respect to the rotation $\mathbf{P}'$, let $\mu^k$ denote the (possibly unstable) match achieved at the end of Phase 1 in the best response dynamics, and let $\mu'$ denote the stable match achieved at the end of Phase 2. The proof proceeds in two parts, each of which establishes a claim regarding Phase 1 and Phase 2, respectively.

First, we claim that every proposer who is matched at the end of Phase 1 weakly prefers their partner in $\mu^k$ to their partner in $\mu^+$. Suppose that proposer $P_x$ is matched in $\mu^k$, i.e., $\mu^k(P_x) \neq P_x$. Recall that in Phase 1, only matched proposers play best responses; unmatched proposers do nothing. Thus, if $P_x$ is matched at the end of Phase 1, then at the least, they weakly prefer their partner in $\mu^k$ to their partner in $\mu^{-i}$. If $P_x \notin \mathbf{P}'$, then the claim is certainly true, since in this case, $\mu^+(P_x) = \mu^{-i}(P_x)$. Suppose instead that $P_x \in \mathbf{P}'$. We claim that if $\mu^+(P_x) \succ_j \mu^k(P_x)$, then at least one proposer is not playing a best response in $\mu^k$. Without loss of generality, let $P_{i_1} = P_x$ and let $A_{j_1} = \mu^+(P_{i_1})$ denote the partner of proposer $P_{i_1}$ in $\mu^+$. Consider the following three cases:
\begin{itemize}[leftmargin=*]
    \item Case 1: $A_{j_1}$ is single or matched with someone they prefer less than $P_{i_1}$ in $\mu^k$. In this case, $P_{i_1}$ is not playing a best response, since they could propose to and match with $A_{i_1}$ instead, whom they prefer to their current partner.
    \item Case 2: $A_{j_1}$ is matched with a proposer $P_y \notin \mathbf{P}'$ whom they prefer to $P_{i_1}$. Since $P_y$ is not in the rotation, it must be the case that $P_y$ is not matched with $A_{j_1}$ in $\mu^+$. This implies that $P_y$ must strictly prefer $A_{j_1}$ to their partner in $\mu^+$, but this would imply that $\mu^+$ is unstable, which is a contradiction.
    \item Case 3: $A_{j_1}$ is matched with a proposer $P_{i_2} \in \mathbf{P}'$ whom they prefer to $P_{i_1}$. By the previous reasoning, $P_{i_2}$ must not prefer $A_{i_1}$ to their partner in $\mu^+$, whom we denote as $A_{j_2}$. The relationship between $P_{i_2}$ and $A_{j_2}$ must also fall in one of the three cases discussed presently. If it falls in Case 1, then $P_{i_2}$ is not playing a best response, and if it falls in Case 2, it would imply a contradiction. Suppose then that it falls into Case 3, so that $A_{j_2}$ prefers $P_{i_3}$, and generalize this idea so that the sequence of proposers $P_{i_1}, P_{i_2}, \dots, P_{i_q}$ fall into Case 3, where each $P_{i_k}$ seeks to be matched with $A_{i_k}$, their partner in $\mu^+$. Because the near-stable match $\mu^{-i}$ was obtained by making proposer $P_i \in \mathbf{P}'$ single, we must have that $q$ is strictly less than $r$. However, by the pigeonhole principle, this would imply that two proposers are matched with the same acceptor in $\mu^+$, which is a contradiction. 
\end{itemize}
These cases establish the claim that if $\mu^+(P_{i_1}) \succ_{i_1} \mu^k(P_{i_1})$, then at least one proposer is not playing a best response in $\mu^k$. In turn, this establishes the first claim that every proposer who is matched at the end of Phase 1 weakly prefers their partner in $\mu^k$ to their partner in $\mu^+$.

Next, we discuss Phase 2. Consider a match $\mu^k$ in which every matched proposer weakly prefers their partner in $\mu^k$ to their partner in $\mu^+$. We claim that if an unmatched proposer $P_x$ plays a best response to $\mu^k$, generating the match $\mu^{k+1}$, then every matched proposer in $\mu^{k+1}$ will weakly prefer their partner in $\mu^{k+1}$ to  their partner in $\mu^+$. This amounts to showing that $P_x$ can at least match with $A_x$, their partner in $\mu^+$. If $A_x$ is unmatched, then $P_x$ can certainly match with $A_x$; otherwise, suppose $A_x$ is matched with $P_y$. It must be the case that $A_x$ prefers $P_x$ to $P_y$; if this were not true, then $P_y$ and $A_x$ would both prefer one another to their partners in $\mu^+$, which would contradict the assumption that $\mu^+$ is stable. Thus, the second claim is established.

After having established these two claims, the proof of Lemma \ref{lemma:lattice_traversal} is essentially complete. Phase 1 of the best response dynamics generates a match $\mu^k$ that every matched proposer weakly prefers to $\mu^+$, and every best response update performed by an unmatched proposer in Phase 2 preserves this property. The number of best response updates in each phase is finite, leading to a stable match which every proposer weakly prefers to $\mu^+$. Importantly, the proof is agnostic to the order in which the best response updates are executed within each Phase.

The proof of Corollary \ref{cor:lattice_stays} follows from the fact that if $P_i \notin \mathbf{P}'$, then Phase 1 will terminate at a match $\mu^k$ that every matched proposer weakly prefers to $\mu$, but not necessarily $\mu^+$. However, the previous arguments regarding Phase 2 are identical, so the best response dynamics will still terminate at a match that every proposer weakly prefers to $\mu$.
\end{proof}

\subsection{Proof}\label{subsection:proof}
In this section, we combine the results from each of the previous sections to complete our proof of Theorem 1. Observe that the learning rule induces a regularly perturbed Markov process over the finite state space
\begin{equation*}
    \mathbf{X} \triangleq \prod_{i = 1}^n \{C, D, W\} \times (\mathbf{A} \cup \{P_i\} \times (\text{ran}(O_{P_i}) \cup \{0\}),
\end{equation*}
where $\text{ran}(f)$ denotes the range of the function $f$. \emph{We must show that the stochastically stable states of this process correspond to the POSM.} We write the probability transition matrix over the state space $\mathbf{X}$ as $\mathbf{T}^\epsilon$ for each $\epsilon > 0$. We must first identify the recurrence classes of the unperturbed process $\mathbf{T}^0$. Let $\mathbf{Z} \subseteq \mathbf{X}$ denote the subset of states in which every agents' benchmark utilities are aligned with their benchmark actions, i.e., no agent is watchful. Let $\mathbf{C}$ denote the subset of states in which at least one matched (content) proposer is not playing a best response, and let $\mathbf{D}$ denote the subset of states in which every matched proposer is playing a best response, but there is at least one unmatched (discontent) proposer. Let $\mathbf{E} \subset \mathbf{Z}$ denote the subset of states whose benchmark action profiles correspond to stable matches, and let $e^* \in \mathbf{E}$ denote the state whose benchmark action profile correspond to the POSM. Observe that $\mathbf{C}$, $\mathbf{D}$, and $\mathbf{E}$ form a partition of $\mathbf{Z}$.

The recurrence classes of the unperturbed process are all singletons $\{x\}$ such that $x \in \mathbf{Z}$. To see this, first consider a state in which every agent is not watchful. Observe that in the unperturbed process, no one experiments, so every content and discontent proposer will repeat their baseline action and receive their same baseline utility. Thus, the process remains in state $x$ with probability one. Next, consider a state in which one more agents are watchful. Once again, every proposer will repeat their baseline action, and after receiving their utility, every watchful proposer will become content or discontent with probability one. Hence their benchmarks will align, meaning that the process arrives at a state in $\mathbf{Z}$ in exactly one transition.

Next, we introduce some important definitions that will be essential for the remainder of our proof. Define the function $r^* : \mathbf{Z} \to \mathbb{R}$ as follows:
\begin{equation*}
r^*(z) \triangleq \min \big\{r(z, z') \big| \; z' \in \mathbf{Z} \setminus \{z\} \big\}.
\end{equation*}
Let $G$ be the graph constructed from $\mathbf{Z}$ as described in Section \ref{subsection:stochastic}. We say that the edge from $z$ to $z'$ is \textit{easy} if $r(z, z') = r^*(z)$. Similarly, we say that a path or tree is easy if each of its constituent edges are easy. The remainder of the proof involves showing that easy paths leads to the POSM, implying that the only easy tree is rooted at the POSM. We begin by identifying the nature of easy transitions for various states.

First, consider a state $z \in \mathbf{C}$, and suppose that the baseline action profile is such that at least one content proposer is not playing a best response. We claim that any easy edge from $z$ to some other state $z'$ must involve a content proposer updating to a best response action. To see this, observe that the only easy transitions out of $z$ require experimentation. In particular, proposers can induce a transition out of $z$ by experimenting twice in succession without updating their baseline actions, causing at least one proposer to first become watchful and then discontent. Every other possible easy transition out of $z$ requires a proposer to experiment and update their baseline action: the resistance of a transition whereby a content proposer chooses to remain single is $2$, the resistance of a transition whereby a discontent proposer chooses to experiment and update their action is $1.5 + F(\Delta u)$, and the resistance of a transition whereby a content proposer chooses to experiment and update their action is $1 + G(\Delta u)$, where $\Delta u$ is the change in utility that the updating proposer experiences. Since $0 \leq F(\cdot), G(\cdot) < 0.5$, it is clear that a transition of minimum resistance must involve a content proposer experimenting and updating their action; moreover, since $G$ is strictly monotone decreasing, this update must be a best response. Given that no one experiments in the following timestep, which occurs with probability $O(1)$, the process transitions to a state $z' \in \mathbf{Z}$ in two timesteps, and the total resistance along this path is $1 + G(\Delta u)$.

Next, consider a state $z \in \mathbf{D}$, and suppose that the baseline action profile is such that every content proposer is playing a best response, and there is at least one discontent proposer. We claim that any easy edge from $z$ to some other state $z'$ must involve a discontent proposer updating to a best response action. To see this, first notice that since content proposers are already playing best responses, they cannot unilaterally deviate to improve their utilities. This implies that any transition out of $z$ in which a content proposer experiments must involve at least two simultaneous or consecutive experimentations, which would have resistance at least $2$. Hence, a transition of minimum resistance must involve a discontent proposer experimenting and updating their action; moreover, since $F$ is strictly monotone decreasing, this update must be a best response. Given that no one experiments in the following timestep, which occurs with probability $O(1)$, the process transitions to a state $z' \in \mathbf{Z}$ in two timesteps, and the total resistance along this path is $1 + F(\Delta u)$, which is strictly less than $2$.

Lastly, consider a state $e \in \mathbf{E}$. We claim that $r^*(e) = 2$, and that any easy transition out of $e$ can only occur in one of two ways. First, observe that since $e$ is a stable match, any unmatched proposer cannot unilaterally deviate to affect the outcome of any agent, including themselves. Hence, an easy transition will not involve a discontent proposer. However, by the same argument as before, a content proposer can induce a transition out of $z$ with resistance $2$. In particular, two consecutive experimentations can cause a transition to a state $z'$ in which the baseline action profile is a near-stable match. However, this form of transition may not be possible depending on the preference structure $\mathbf{O}$. The other form of easy transition out of $z$ with resistance $2$ involves a content proposer choosing to remain single, which causes a one-timestep transition to a state $z'$ in which the baseline action profile is a near-stable match. This form of transition is always possible.

The preceding arguments imply that every easy path starting from a state $z \notin \mathbf{E}$ consists of a sequence of baseline action profiles which are generated by the best response dynamics. In particular, an easy transition out of a state in $\mathbf{C}$ involves a matched proposer playing a best response, which occurs in Phase 1 of the best response dynamics, and an easy transition out of a state in $\mathbf{D}$ involves an unmatched proposer playing a best response, which occurs in Phase 2 of the best response dynamics.

Having established the nature of these easy paths, we can now provide a simple construction of an easy tree rooted at $e^*$. Initialize $T(e^*)$ as an empty set of edges, and add to it as follows. Consider a vertex $z_1 \in \mathbf{Z} \setminus \mathbf{E}$. If $z_1$ is not part of an edge already in $T(e^*)$, then add the easy edge from $z_1$ to $z_2$. Continue adding edges in this fashion, traversing nodes $z_3, \dots, z_{k}$ until either $z_{k}$ already has an outgoing edge in $T(e^*)$, or until reaching an equilibrium $z_{k} = e_1 \in \mathbf{E}$. Repeat this process until every node in $\mathbf{Z} \setminus \mathbf{E}$ has exactly one outgoing edge. Then, for every vertex $e \in \mathbf{E} \setminus \{e^*\}$, add an easy edge from $e$ to a state $z \in \mathbf{Z}$ whose baseline action constitutes a near-stable match; in particular, choose a state such that the proposer made single belongs to a rotation with respect to the stable match represented by $e_1$. By Lemma \ref{lemma:lattice_traversal}, the easy path from $e$ will lead to an equilibrium that every proposer weakly prefers to a next-stable match with respect to the stable match given by $e$, because a proposer in a rotation is made single. It is straightforward to show that the properties of the best response dynamics ensure that this construction does not introduce any cycles. Thus, there exists a unique easy path from every vertex $z \in \mathbf{Z} \setminus \{e^*\}$ to $e^*$ in $T(e^*)$, so $T(e^*)$ is indeed an easy tree rooted at $e^*$.

All that remains to be shown is that a minimum resistance tree rooted at any other node $z \neq e^*$ has resistance strictly greater than $\rho(e^*)$, the stochastic potential of $e^*$. First, consider any $z \in \mathbf{C} \cup \mathbf{D}$, and let $T(z)$ be a minimum resistance tree rooted at $z$. Every vertex $z' \in \mathbf{Z} \setminus \{z, e^*\}$ contributes resistance at least $r^*(z')$, the vertex $e^*$ contributes resistance at least $2$, and the resistance corresponding to $z$ is removed, so the stochastic potential of $T(z)$ is at least $\rho(z) \geq \rho(e^*) + 2 - r^*(z)$. Because $z$ is not an equilibrium, $r^*(z)$ is strictly less than $2$, so $\rho(z) > \rho(e^*)$. Next, consider a state $e \in \mathbf{E}$. Since $r^*(e) = r^*(e^*) = 2$, we have that $\rho(e) \geq \rho(e^*)$. Because $T(e)$ is a tree rooted at $e$, there exists a unique path from $e$ to $e^*$. However, Corollary \ref{cor:lattice_stays} implies that every best response sequence starting from a near-stable match with respect to $\mu^*$ must lead back to $\mu^*$. Thus, the path from $e$ to $e^*$ in $T(e)$ must include at least one transition, say, from $z$ to $z'$, which is not a best response update (if $z \in \mathbf{C} \cup \mathbf{D}$) or does not lead to a near-stable match (if $z \in \mathbf{E}$). This transition is not easy, so it has resistance strictly greater than $r^*(z)$; thus, $\rho(e) > \rho(e^*)$. This establishes that the unique recurrence class of minimum stochastic potential is $e^*$, and hence, the POSM is the only stochastically stable state.


\bibliographystyle{ieeetr}
\bibliography{references}

\end{document}